\documentclass{ws-ijmpa}

\usepackage{dcolumn}
\usepackage{amsmath}
\usepackage{colordvi}

\newcommand{\eV}{\mbox{\rm eV}}
\newcommand{\keV}{\mbox{\rm keV}}
\newcommand{\MeV}{\mbox{\rm MeV}}
\newcommand{\GeV}{\mbox{\rm GeV}}

\renewcommand{\draftnote}{}

\begin{document}

\markboth{D. Kekez, D. Klabu\v{c}ar and M. D. Scadron}
{Bypassing the axial anomalies}

%
%

\title{BYPASSING THE AXIAL ANOMALIES}

\author{DALIBOR KEKEZ}

\address{Rudjer Bo\v{s}kovi\'{c} Institute,
	P.O.B. 180, 10002 Zagreb, Croatia \\ kekez@lei.irb.hr}
 
\author{DUBRAVKO KLABU\v{C}AR\footnote{Senior Associate of Abdus Salam ICTP}}

\address{Physics Department, Faculty of Science, University of Zagreb \\
	P.O.B. 331, 10002 Zagreb, Croatia \\
	klabucar@oberon.phy.hr}
 
\author{M. D. SCADRON}

\address{Physics Department, University of Arizona \\
	Tucson Az 85721 USA \\
	scadron@physics.arizona.edu}

\maketitle

\begin{abstract}
Many meson processes are related to the $U_A(1)$
axial anomaly, present in the Feynman graphs where
fermion loops connect axial vertices with vector
vertices.
However, the coupling of pseudoscalar mesons to quarks
does not have to be formulated via axial vertices.
The pseudoscalar coupling is also possible, and this
approach is especially natural on the level of
the quark substructure of hadrons.
In this paper we point out the advantages of calculating
these processes using (instead of the anomalous graphs)
the Feynman graphs where axial vertices are replaced by
pseudoscalar vertices.  We elaborate
especially the case of the processes related to the
Abelian axial anomaly of QED, but we speculate that
it seems possible that effects of the non-Abelian
axial anomaly of QCD can be accounted for in an
analogous way.

\end{abstract}

\ccode{PACS numbers: 14.40 -n, 12.39.Fe, 13.20.-v, 11.10.St}

\section{Introduction}

\begin{figure}
\centerline{\includegraphics[height=40mm,angle=0]{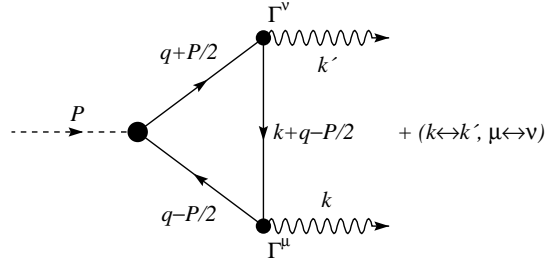}}
\caption{The triangle graph and its crossed graph 
relevant for the interaction of the neutral pseudoscalar meson
of momentum $P$ with two photons of momenta $k$ and $k^\prime$.
The quark-photon coupling is in general given by dressed vector
vertices $\Gamma_\mu(q_1,q_2)$, which in the free limit reduce 
to $e {\cal Q} \gamma_\mu$.}
\label{fig:triangle}
\end{figure}

\begin{figure}
\centerline{\includegraphics[height=40mm,angle=0]{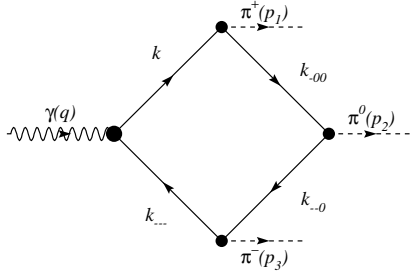}}
\caption{One of the box diagrams for the process
$\gamma \to \pi^+ \pi^0 \pi^-$, studied by the pseudoscalar coupling 
method in, e.g.,
                 Refs.~3, 4. 
There are six different contributing graphs, obtained from
the above graph
by the permutations of the vertices of the three different pions.
The position of the $u$ and $d$
quark flavors on the internal lines, as well as $Q_u$ or $Q_d$
quark charges in the quark-photon vertex, varies from graph to
graph, depending on the position of the quark-pion vertices.
The physical pion fields are $\pi^\pm=(\pi_1\mp i \pi_2)/\sqrt{2}$
and $\pi^0\equiv\pi_3$. 
The momenta flowing through the four sections of the quark loop
are conveniently given by various combinations of the symbols
$\alpha, \beta, \gamma = +, 0, -$ in $k_{\alpha\beta\gamma} \equiv 
k + \alpha p_{1} + \beta p_{2} + \gamma p_{3}$.}
\label{fig:box}
\end{figure}

Numerous processes in meson physics are related to the 
Adler-Bell-Jackiw (ABJ) axial anomaly \cite{Adler69,BellJackiw69}
appearing in the fermion loops connecting certain number of axial (A) 
and vector (V) vertices. 
Concretely, in this paper we will deal with the processes
related to the AVV (``triangle", Fig.~\ref{fig:triangle}) 
and VAAA (``box", Fig.~\ref{fig:box}) 
anomaly, exemplified by the famous $\pi^0 \to \gamma\gamma$ and
$\gamma \to \pi^+ \pi^0 \pi^-$ transitions. 

Suppose one wants to describe such processes using QCD-related
effective chiral meson Lagrangians \cite{seeGeorgi,Georgi:1985kw}
without adding ad hoc interactions of mesons with external
gauge fields to reproduce empirical results. For example, one
can add by hand 
\begin{equation}
\Delta {\cal L} = g_{\pi\gamma\gamma} \pi^0
\epsilon_{\mu\nu\rho\sigma} F^{\mu\nu} F^{\rho\sigma} \, ,
\label{L_pi2gamma}
\end{equation}
and this would reproduce the observed $\pi^0 \to \gamma\gamma$
width for the favorable value of the $\pi^0\gamma\gamma$
coupling $g_{\pi\gamma\gamma}$. However, if one does not want 
to add such ad hoc terms in the effective meson Lagrangians,
one  must describe such ``anomalous'' processes through the term
derived by Wess and Zumino \cite{WZ}.  On the other hand, if one 
wants to utilize and explicitly take into account the fact that 
mesons are composed of quarks, another way of describing these
processes is optimal in our opinion, and the main purpose of this 
paper is to stress and elucidate this. 

Axial vertices in the anomalous graphs such as the AVV and VAAA ones,
couple the quarks with pseudoscalar mesons. 
Instead of anomalous graphs, another way to study the related 
amplitudes involving pseudoscalar mesons, is to calculate the 
corresponding graphs where axial vertices (A) are replaced by 
pseudoscalar (P) ones. Thereby, for example, the 
$\pi^0 \to \gamma\gamma$
decay amplitude due to the AVV ``triangle anomaly",
\begin{equation}
F_{m_\pi=0}(\pi^0\to 2\gamma)
                           = \frac{e^2 N_c}{12\pi^2f_\pi}~,
\label{pi2gammaAmp}
\end{equation}
is reproduced by the calculation of the PVV triangle graph. 
[Eq. (\ref{pi2gammaAmp}) pertains to the chiral limit, where 
the pion mass $m_\pi=0$. Also, $f_\pi\approx 93~\MeV$
is the pion decay constant, 
$e$ is the proton charge, and $N_c=3$ is the number of quark colors.]

The PVV triangle graph calculation of Eq. (\ref{pi2gammaAmp})
can most simply be done essentially {\it {\` a} la} Steinberger 
\cite{S49}, that is, 
with a loop of ``free'' constituent quarks with the point 
pseudoscalar coupling (i.e., $g \gamma_5$, where $g = constant$) 
to quasi-elementary pion fields.
However, since the development of the Dyson-Schwinger (DS) 
approach to quark-hadron physics \cite{Alkofer:2000wg,Maris:2003vk}, 
the presently advocated method becomes even more convincing. 
Namely, the DS approach clearly shows how the light pseudoscalar 
mesons simultaneously appear both as quark-antiquark ($q\bar q$) 
bound states and as Goldstone bosons of the dynamical chiral 
symmetry breaking (D$\chi$SB) of nonperturbative QCD. The 
solutions of Bethe-Salpeter (BS) equations for the bound-state 
vertices of pseudoscalar mesons then enter in the PVV triangle 
graph instead of the point $g \gamma_5$ coupling, and the 
current algebra result (\ref{pi2gammaAmp}) is again reproduced 
exactly and analytically, which is unique among the bound-state
approaches. That the (almost massless) pseudoscalars are 
(quasi-)Goldstone bosons, is also a unique feature among 
the bound-state approaches to mesons.
   
A reason why the P-coupling method is simpler both technically 
and conceptually is that the PVV triangle graph amplitude is 
{\em finite}, unlike the AVV one, which is divergent and therefore 
also ambiguous with respect to the momentum routing.
Reference \refcite{Kekez:2005kx} gives a more detailed discussion of 
the P-coupling method and why is that it is equivalent to the 
anomaly calculations, and illustrates this on
the examples of the famous decay $\pi^0 \to \gamma \gamma$
and processes of the type $\gamma \to \pi^+ \pi^0 \pi^-$.
Also, the PVV quark triangle amplitude leads to many (over 15)
decay amplitudes in agreement with data to within 3\% and not
involving free parameters \cite{Delbourgo,Delbourgo:1993dk,Bramon:1997gg}. 
This will be elaborated 
in more detail in Sec. \ref{manyProc}. Additional advantages
of this method is that its treatment of the $\eta$-$\eta'$ complex 
and resolution of the $\rm U_A(1)$ problem, goes well with the 
absence of axions (which were predicted to solve the strong CP 
problem but have {\it not} yet been observed \cite{PDG2004}) and 
with the arguments of Ref.~\refcite{Banerjee:2000qw}, that there is really 
no strong CP problem. All this will be discussed in
Sec.~\ref{sec:CommRelToGlounAnom}.
We state our conclusions in Sec.~\ref{sec:summary}.

\section{Processes Going Through the Quark Triangle}
\label{manyProc}

\begin{figure}
\begin{center}
\begin{tabular}{cc}
\includegraphics[height=26mm,angle=0]{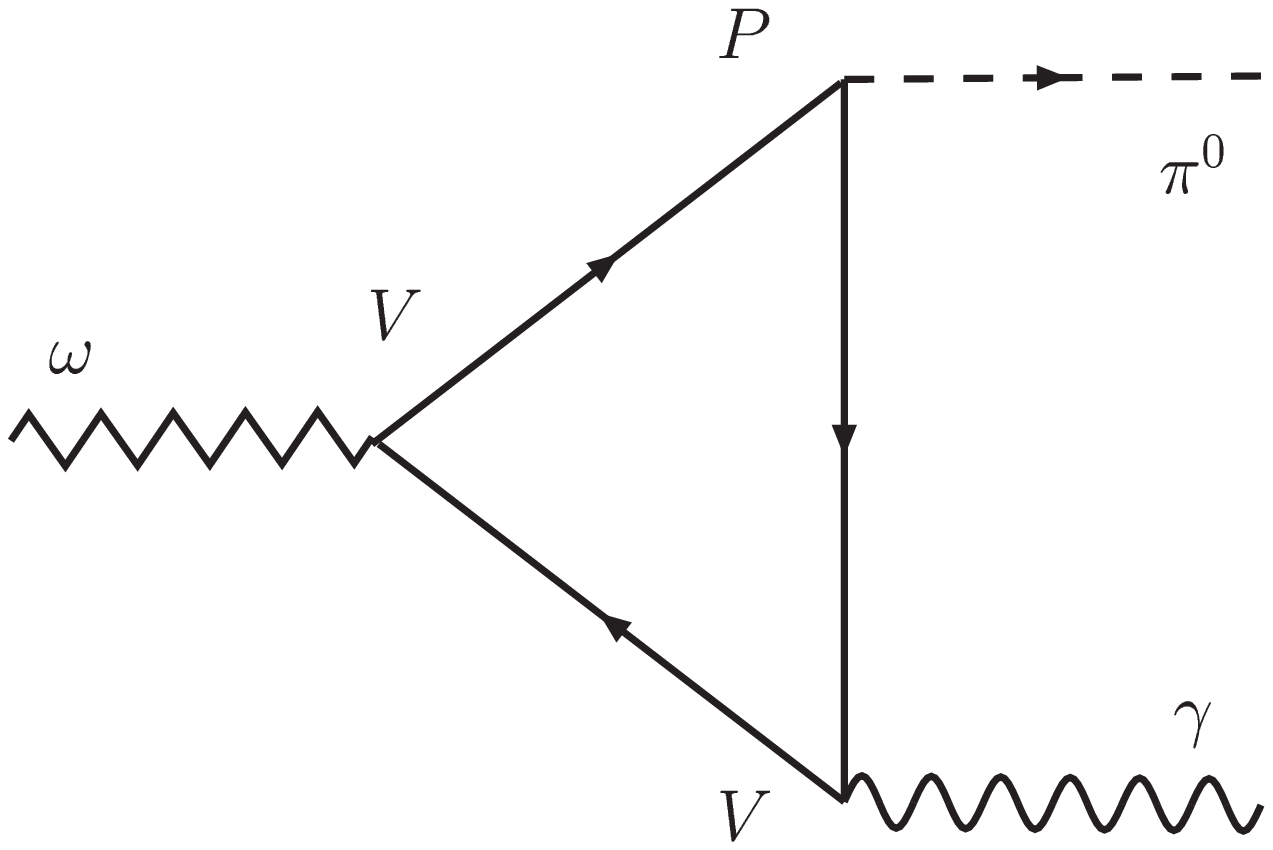}
&
\includegraphics[height=26mm,angle=0]{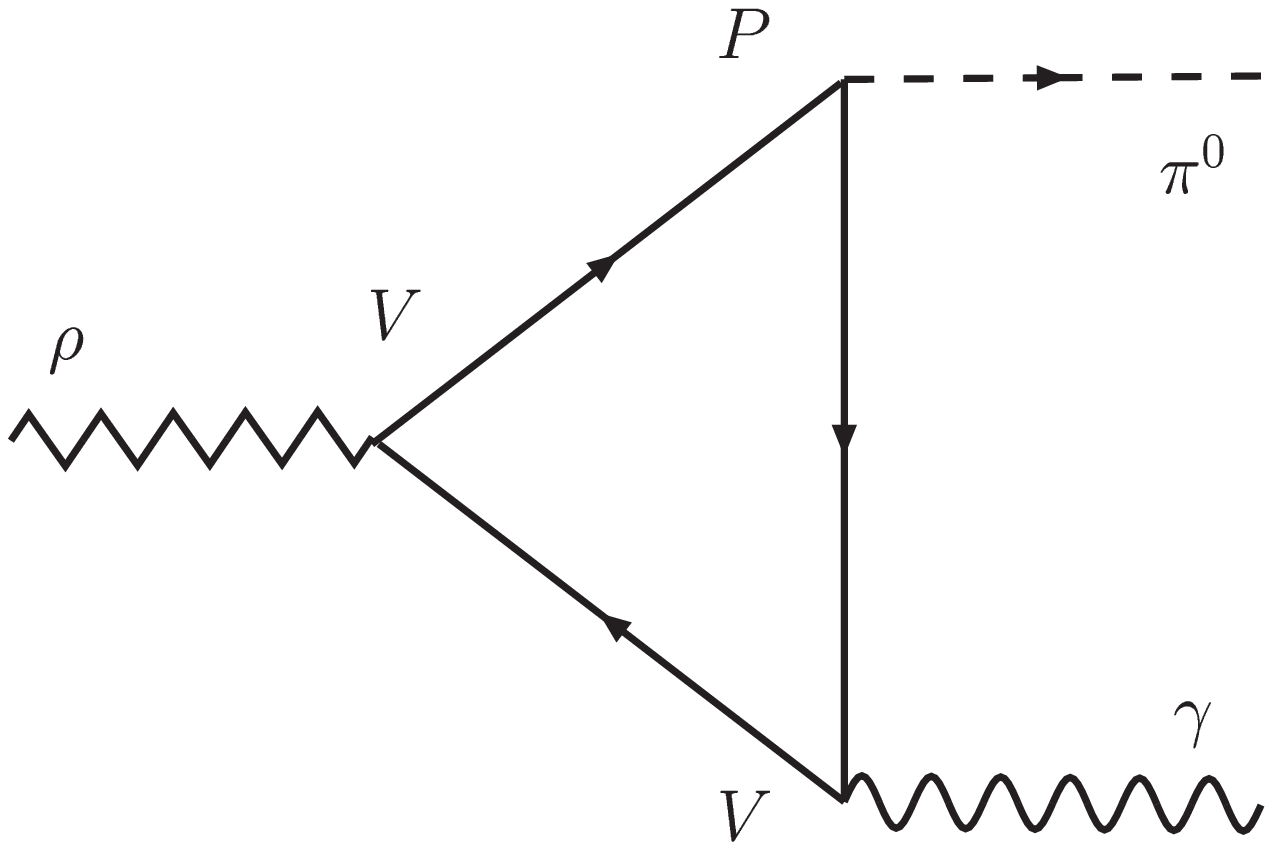}
\end{tabular}
\end{center}
\caption{Two examples of the PVV triangle graphs
where just one of the vector vertices couples to a photon,
whereas the other couples to a vector meson. These two graphs
describe the decays of $\omega$ and $\rho$ mesons into a photon
and a pion.}
\label{fig:PVVdiagrams}
\end{figure}

 In this section we calculate the amplitudes for a number of processes
using the quark triangle graphs.
Figures \ref{fig:triangle} and 
\ref{fig:PVVdiagrams} show three such PVV processes.
First we consider $\pi^0\to \gamma\gamma$ decay via the $u$ and $d$ 
quark triangle graph for $\pi^0=(\bar{u}u-\bar{d}d)/\sqrt{2}$, $N_c=3$ 
and Goldberger-Trieman relation 
leading to the pion decay constant: 
$f_\pi=\hat{m}/g_{\pi qq}$.
This amplitude is finite and for the experimental value of the pion 
decay constant, $f_\pi = (92.42\pm 0.26) \, \MeV$ \cite{PDG2004},
gives \cite{Delbourgo} the chiral-limit amplitude (\ref{pi2gammaAmp})
of magnitude

\begin{equation}
|F_{m_\pi=0}(\pi^0\to 2\gamma)|
=
\frac{e^2}{4\pi^2 f_\pi}=0.0251~\GeV^{-1}
\end{equation}

\noindent very close to experimental data \cite{PDG2004}

\begin{equation}
|F_{\rm exp}(\pi^0\to 2\gamma)| =
\left[\frac{64\pi\Gamma(\pi^0\to\gamma\gamma)}{m_\pi^3}\right]^{1/2}
=
(0.0252\pm 0.0009)~\GeV^{-1}~.
\end{equation}

\noindent Likewise, the $u$, $d$ quark triangles
for $\rho\to\pi\gamma$ decay give \cite{Delbourgo}

\begin{equation}
|F(\rho\to\pi\gamma)|=\frac{e g_\rho}{8\pi^2 f_\pi}
= 0.206~\GeV^{-1}
\end{equation} 

\noindent for $g_\rho=4.965\pm 0.002$
found from $\rho^0\to e^-e^+$ decay
\cite{PDG2004}:

\begin{equation}
\Gamma(\rho^0\to e^-e^+)
=
\frac{e^4 m_\rho}{12\pi g_\rho^2}
=
(7.02\pm 0.11)~\keV~.
\end{equation}

\noindent The calculated $|F(\rho\to\pi\gamma)|$
is also near data 
\cite{PDG2004},
\begin{equation}
|F_{\rm exp}(\rho\to\pi\gamma)|
=
\left[\frac{12\pi\Gamma(\rho\to\pi\gamma)}{q^3}\right]^{1/2}
=(0.225\pm 0.011)~\GeV^{-1}~,
\end{equation}
where $q=(m_\rho^2-m_\pi^2)/(2m_\rho)$ is the photon momentum.
[Actually, the above value is a weighted average of
$F_{\rm exp}(\rho^0\to\pi^0\gamma)$
and $F_{\rm exp}(\rho^\pm\to\pi^\pm\gamma)$ amplitudes.]

Next we predict
the $u$, $d$ quark triangle amplitude for $\omega\to\pi\gamma$
taking $\omega$
as 99\% nonstrange \cite{PDG2004} ($\cos^2\phi_V \approx 0.99$)

\begin{equation}
|F(\omega\to\pi\gamma)|
=
\frac{\cos\phi_V\, e\, g_\omega}{8\pi^2 f_\pi} = 0.705~\GeV^{-1}
\label{FOmegaToPi0Gamma}
\end{equation} 

\noindent for $g_\omega=17.06\pm 0.28$
found from $\omega\to e^-e^+$ decay.
The mixing angle is\footnote{We use quadratic mass formulae
for mesons (See, e.g., Ref.~\refcite{PDG2002} and earlier).
However, the input experimental meson masses are newest, 
taken from Ref.~\refcite{PDG2004}.}

\begin{eqnarray}
\phi_V
&=&
\theta_V - \arctan(\frac{1}{\sqrt{2}}) =
\arctan\sqrt{\frac{\frac{1}{3}(4 m_{K^\star}^2 - m_\rho^2) - m_\varphi^2 }
                   {m_\omega^2-\frac{1}{3}(4 m_{K^\star}^2 - m_\rho^2) } }
-\arctan(\frac{1}{\sqrt{2}})
\nonumber \\
&=& (5.208\pm 0.092)^\circ~.
\label{phiV}
\end{eqnarray}

\noindent Again this theory in Eq.~(\ref{FOmegaToPi0Gamma}) is near data
$(0.722\pm 0.012)~\GeV^{-1}$
\cite{PDG2004}.

   Other PVV photon decays involve the $\eta$ and $\eta^\prime$
mixed non--strange and $\bar{s}s$ pseudoscalar mesons. Again the
quark triangle amplitudes are a close match with data
\cite{Delbourgo,Delbourgo:1993dk,Bramon:1997gg}.

  The quark--triangle (QT) calculation gives reliable predictions
also for the $\eta$ and $\eta^\prime$ two--photon decays:

\begin{eqnarray}
|F(\eta\to\gamma\gamma)|
=
\frac{e^2}{4\pi^2 f_\pi}
\frac{N_c}{9}
(5 \cos\phi_P - \sqrt{2} \frac{\hat{m}}{m_s}\sin\phi_P)
=
0.0255\GeV^{-1}~,
\\
|F(\eta^\prime\to\gamma\gamma)|
=
\frac{e^2}{4\pi^2 f_\pi}
\frac{N_c}{9}
(5 \sin\phi_P + \sqrt{2} \frac{\hat{m}}{m_s}\cos\phi_P)
=
0.0345~\GeV^{-1}~.
\end{eqnarray}

\noindent This should be compared with the experimental data:

\begin{eqnarray}
|F_{\rm exp}(\eta\to\gamma\gamma)|
=
\left[\frac{64\pi\Gamma(\eta\to\gamma\gamma)}{m_\eta^3}\right]^{1/2}
=(0.02498\pm 0.00064)~\GeV^{-1}~,
\\
|F_{\rm exp}(\eta^\prime\to\gamma\gamma)|
=
\left[\frac{64\pi\Gamma(\eta^\prime\to\gamma\gamma)}{m_{\eta^\prime}^3}\right]^{1/2}
=(0.03133\pm 0.00055)~\GeV^{-1}~,
\end{eqnarray}

\noindent where $\Gamma(\eta\to\gamma\gamma)
=(0.5108\pm 0.0268)~\keV$
and 
$\Gamma(\eta^\prime\to\gamma\gamma)
=(4.29\pm 0.15)~\keV$. The ratio of the constituent quark masses
is $m_s/m=2f_K/f_\pi-1=1.445\pm 0.024$
for $f_{\pi^\pm}=(92.4\pm 0.3)~\MeV$
and $f_K=(113.0\pm 1.0)~\MeV$ \cite{PDG2004}.
The mixing angle is \cite{Jones:1979ez,Kekez:2000aw}

\begin{eqnarray}
\phi_P &=& 
\theta_P + \arctan({\sqrt{2}}) =
\arctan
        \sqrt{
                \frac   {(m_{\eta'}^2 - 2m_K^2 + m_\pi^2) (m_\eta^2 -
                                 m_\pi^2)}
                                {(2m_K^2 - m_\pi^2 - m_\eta^2) (m_{\eta'}^2
                                 - m_\pi^2)}
        }
\nonumber \\
&=& (42.441\pm 0.019)^\circ~.
\label{phiP}
\end{eqnarray}

   Next, we can calculate the $\rho^0\to\eta\gamma$ amplitude employing
the quark--triangle diagram,

\begin{equation}
|F(\rho^0\to\eta\gamma)|
=
\frac{e g_\rho}{8\pi^2 f_\pi} 3\cos\phi_P
=0.456~\GeV^{-1}~.
\end{equation}

\noindent Again, this is close to the experimental data,

\begin{equation}
|F_{\rm exp}(\rho^0\to\eta\gamma)|
=
\left[\frac{12\pi\Gamma(\rho^0\to\eta\gamma)}{q^3}\right]^{1/2}
=(0.48\pm 0.03)~\GeV^{-1}~,
\end{equation}

\noindent where
$q=(m_\rho^2-m_\eta^2)/(2m_\rho)=(194.5\pm 0.4)~\MeV$
is the photon momentum and
$\Gamma(\rho^0\to\eta\gamma)
=(45.1\pm 6.0)~\keV$.
A similar situation is with the $\eta^\prime\to\rho\gamma$ amplitude,
for which the quark--triangle calculation gives

\begin{equation}
|F(\eta^\prime\to\rho^0\gamma)|
=
\frac{e g_\rho}{8\pi^2 f_\pi} 3\sin\phi_P
=
0.417~\GeV^{-1}~.
\end{equation}

\noindent The corresponding experimental value is

\begin{equation}
|F_{\rm exp}(\eta^\prime\to\rho^0\gamma)|
=
\left[\frac{4\pi\Gamma(\eta^\prime\to\rho^0\gamma)}{q^3}\right]^{1/2}
=(0.411\pm 0.017)~\GeV^{-1}~,
\end{equation}

\noindent where
$q=(m_{\eta^\prime}^2-m_\rho^2)/(2m_{\eta^\prime})
=(164.7\pm 0.4)~\MeV$ is the photon momentum and

\begin{equation}
\Gamma(\eta^\prime\to\rho^0\gamma\,\,\,\mbox{\rm including non--resonant}\,\,\, \pi^+\pi^-\gamma)
=(60.0\pm 5.0)~\keV
\end{equation}

\noindent is the experimental decay width \cite{PDG2004}.

   The $\eta\to\pi\pi\gamma$ amplitude is

\begin{equation}
|M^{\mbox{\rm\scriptsize VMD}}_{\eta\to\pi\pi\gamma}|
=
|\frac{2g_{\rho\pi\pi} M^{\mbox{\rm\scriptsize QT}}_{\rho^0\to\eta\gamma}}
      {m_\rho^2-s}|
=9.80~\GeV^{-3}
\end{equation}

\noindent where $s=m_\pi^2$. The $\eta\to\pi\pi\gamma$ decay width is

\begin{equation}
\Gamma(\eta\to\pi\pi\gamma)
=
\frac{|M_{\eta\to\pi\pi\gamma}|^2}{(2\pi)^3}
m_\eta^{7} Y_\eta
=56.2~\eV~,
\end{equation}

\noindent where $Y_\eta = 0.98\cdot 10^{-5}$ \cite{thew}.
This is in a good agreement with the experimental value

\begin{equation}
\Gamma(\eta\to\pi\pi\gamma)
=(60.4\pm 3.6)~\eV~,
\end{equation}
\noindent revealing that the vector meson dominance 
is the main effect, while the coupling through VPPP quark box loop
(``contact term") contributes little.

  It is known that $\omega\to 3\pi$ decay is dominated by
$\rho$--meson poles. The required $\omega\to\rho\pi$ amplitude can be estimated
as

\begin{equation}
|M^{\mbox{\rm\scriptsize VMD}}(\omega\to\rho\pi)|
=
\left(\frac{g_\rho}{e}\right)
|F(\omega\to\pi^0\gamma)|
\sim 12~\GeV^{-1}~,
\end{equation}

\noindent but cannot be measured because there is no phase space for this
process. The $\omega\to\rho\pi$ amplitude is more precisely defined with
QL, additionally enhanced with a meson loop associated with sigma exchange
\cite{Delbourgo:1993dk,Bramon:1997gg,freund-nandi-rudaz},

\begin{equation}
|M(\omega\to\rho\pi)|_{\mbox{\rm\scriptsize QT}}
=
\frac{3 g_{\rho\pi\pi}^2}{8\pi^2 f_\pi}
\approx
15~\GeV^{-1}~.
\end{equation}

\noindent {The scalar amplitude
$M^{\mbox{\rm\scriptsize VMD}}(\omega\to 3\pi)$
is dominated by the $\rho$ meson in each of the three possible channels
\cite{gell-mann},

\begin{eqnarray}
|M^{\mbox{\rm\scriptsize VMD}}(\omega\to 3\pi)|
&=&
2 g_{\rho\pi\pi}
|M(\omega\to\rho\pi)|
\left[
\frac{1}{m_\rho^2-s}
+
\frac{1}{m_\rho^2-t}
+
\frac{1}{m_\rho^2-u}
\right]
\nonumber \\
&\approx& 1480~\GeV^{-3}~.
\end{eqnarray}

\noindent Following Thew's phase space analysis \cite{thew}, we get

\begin{equation}
\Gamma(\omega\to\ 3\pi)
=
\frac{|M^{\mbox{\rm\scriptsize VMD}}(\omega\to 3\pi)|^2}{(2\pi)^3}
m_\omega^7
Y_\omega
=
7.3~\MeV
\end{equation}

\noindent where $Y_\omega = 4.57\cdot 10^{-6}$ is used.
The predicted value is close to the experimental value \cite{PDG2004}

\begin{equation}
\Gamma(\omega\to 3\pi)
=
(7.6\pm 0.1)~\MeV~.
\end{equation}

Here we have taken $\omega$ as pure NS, although it is about 99\% NS,
since $\phi_V=(5.208 \pm 0.092)^\circ$ from our Eq.~(\ref{phiV}).

   In the quark--level $\sigma$--model a quark box diagram contributes
to the $\omega\to 3\pi$ decay. This box diagram can be interpreted as a
contact term. It is shown that the contact contribution is small by itself,
but can be enlarged through the interference effect
\cite{Lucio-Martinez:2000ea}.

   Using $\phi_P=(42.441\pm 0.019)^\circ$ from our Eq.~(\ref{phiP}),
we predict the tensor $T\to PP$ branching ratios for
$a_2(1320)$:

\begin{eqnarray}
\begin{array}{ll}
\mbox{\rm BR}(\frac{\textstyle{a_2\to\eta\pi}}{\textstyle{a_2\to K\bar{K}}})
=\left(\frac{\textstyle{p_{\eta\pi}}}{\textstyle{p_{K}}}\right)^5
   2 \cos^2\phi_P=2.996
& (\mbox{\rm data}\,\,\,\,
2.96\pm 0.54)~,
\\
\mbox{\rm BR}(\frac{\textstyle{a_2\to\eta^\prime\pi}}
		   {\textstyle{a_2\to K\bar{K}}})
=\left(\frac{\textstyle{p_{\eta^\prime\pi}}}{\textstyle{p_{K}}}\right)^5
   2 \sin^2\phi_P=0.1113
& (\mbox{\rm data}\,\,\,\,
0.108\pm 0.025)~,
\\
\mbox{\rm BR}(\frac{\textstyle{a_2\to\eta^\prime\pi}}
                   {\textstyle{a_2\to\eta\pi}})
=\left(\frac{\textstyle{p_{\eta^\prime\pi}}}{\textstyle{p_{\eta\pi}}}\right)^5 \tan^2 \phi_P=0.0371
& (\mbox{\rm data}\,\,\,\,
0.0366\pm 0.0069)~,
\end{array}
\end{eqnarray}

\noindent for center of mass  momenta
$p_{\eta\pi}=535~\MeV$,
$p_{\eta^\prime\pi}=287~\MeV$,
$p_{K}=437~\MeV$.
The above data branching ratios follow from $a_2(1320)$ recent fractions
\cite{PDG2004}:
$\mbox{\rm BR}(a_2\to\eta\pi)       =(14.5\pm 1.2)\%$,
$\mbox{\rm BR}(a_2\to K\bar{K})     =( 4.9\pm 0.8)\%$ and
$\mbox{\rm BR}(a_2\to\eta^\prime\pi)=( 5.3\pm 0.9)\cdot 10^{-3}$.

\section{Comments Related to the Gluon Anomaly}
\label{sec:CommRelToGlounAnom}}

The approach using the pseudoscalar coupling is, in our opinion,
also relevant for the effects related to the non-Abelian, ``gluon"
ABJ axial anomaly. Here, we comment on this only briefly,
and direct the reader to the original references for details.

\subsection{Goldstone structure and $\eta$-$\eta'$ phenomenology}

The first point concerns the $\eta$-$\eta'$ complex and the $U_A(1)$ 
problem related to it. 

In the chiral limit $m_\pi = m_K = m_{\eta_8} = 0$,
since all members of the flavor-SU(3) pseudoscalar meson octet 
are massless in this theoretical, but very useful limit. The only 
non-vanishing ground-state pseudoscalar meson mass in this limit 
is the mass of the SU(3)-singlet pseudoscalar meson $\eta_1$.
This is thanks to the non-Abelian, gluon ABJ axial anomaly, i.e.,
to the fact that the divergence of the SU(3)-singlet axial current 
\begin{equation}
A^\mu_0(x) = \overline\Psi(x) \gamma^\mu \gamma_5 \Psi(x) \, ,
\end{equation}
receives the contributions from both the finite quark masses
and gluon fields $G_a^{\mu\nu}$:
\begin{equation}
\partial_\mu A^\mu_0 = 
2 {\rm i} m_u \, \overline{u} \gamma_5 u +
2 {\rm i} m_d \, \overline{d} \gamma_5 d +
2 {\rm i} m_s \, \overline{s} \gamma_5 s +
\frac{3 \, g^2 }{32 \pi^2} 
\epsilon_{\mu\nu\alpha\beta} G_a^{\mu\nu} G_a^{\alpha\beta}\, .
\label{divGanom}
\end{equation}
This removes the $U_A(1)$ symmetry and explains why only eight 
pseudoscalar mesons are light, and not nine; i.e., why there
is an octet of (almost-)Goldstone bosons, but not a nonet.
The physically observed $\eta$ and $\eta'$ are then the 
mixtures of the anomalously heavy $\eta_1$ and 
(almost-)Goldstone $\eta_8$ in such a way that $\eta'$ is
predominantly $\eta_1$ and $\eta$ is predominantly $\eta_8$.
This is how the gluon anomaly can save us from the $U_A(1)$ problem
in principle, and the details of how we achieve a successful 
description of the $\eta$-$\eta'$ complex, are given in Refs. 
\refcite{Jones:1979ez,Kekez:2000aw,Klabucar:1997zi,Klabucar:2000me,Klabucar:2001gr}. 
Here we just sketch some important points.
The mass matrix squared $\hat{M}^2$
in the quark basis $|u\bar{u}\rangle$, $|d\bar{d}\rangle$, $|s\bar{s}\rangle$ 
is

\begin{equation}
\hat{M}^2 
=
\hat{M}_{\mbox{\rm\scriptsize NA}}^2 + \hat{M}_{\rm A}^2
=
\left[
\begin{array}{ccc}
m_{u\bar{u}}^2 & 0 & 0 \\
0 & m_{d\bar{d}}^2 & 0 \\
0 & 0 & m_{s\bar{s}}^2
\end{array}
\right]
+
\beta
        \left[ \begin{array}{ccl} 1 & 1 & X \\
                                  1 & 1 & X \\
                                  X & X & X^2
        \end{array} \right]~,
\end{equation}

\noindent 
where $\hat{M}_{\mbox{\rm\scriptsize NA}}^2$ is the non-anomalous part of the matrix,
since $m_{u\bar{u}}^2=m_{d\bar{d}}^2=m_\pi^2$ and
$m_{s\bar{s}}^2=2m_K^2-m_\pi^2$ would be the masses of the respective
``non-strange" (NS) and ``strange" (S) pseudoscalar $q\bar{q}$ mesons
if there were no gluon anomaly.
In the NS sector, in the isospin symmetry limit 
(which is very close to reality), the relevant combinations are 
$| \pi^0 \rangle = | u\bar{u} - d\bar{d} \rangle / \sqrt{2}$
as the neutral partner of the charged pions $| \pi^\pm \rangle$
in the isospin 1 triplet, and the isospin 0 combination
$ | u\bar{u} + d\bar{d} \rangle / \sqrt{2} $.
In the absence of gluon anomaly, but with
an $s$-quark mass heavier than the isosymmetric $u$ and $d$ ones, 
$\eta$ would reduce to $|{\rm NS}\rangle=|u\bar{u} + d\bar{d} \rangle / \sqrt{2}$
with the mass $m_{\mbox{\rm\scriptsize NS}} = m_\pi$,
and $\eta'$ to $|{\rm S}\rangle = | s\bar{s} \rangle $ with the mass
$m_{\mbox{\rm\scriptsize S}} = m_{s\bar{s}}$. Both of these assignments are in conflict
with experiment.  The realistic contributions of various flavors to
$\eta$ and $\eta'$ and their masses (i.e., the realistic
$\eta$-$\eta'$ mixing) are obtained only thanks to $\hat{M}_{\rm A}^2$,
the anomalous contribution to the mass matrix. In $\hat{M}_{\rm A}^2$,
the quantity $\beta$ describes transitions
$|q\bar{q}\rangle\to|q^\prime\bar{q}^\prime\rangle$
($q,q^\prime=u,d,s$) due to the gluon anomaly and $X$
describes the effects of the SU(3) flavor symmetry breaking
on these transitions.
In Refs.~\refcite{Jones:1979ez,Kekez:2000aw,Klabucar:2000me},
as the first step 
in solving the $U_A(1)$ problem,
we extract $\eta_8$, $\eta_1$ masses from the $\eta$, $\eta^\prime$
via

\begin{eqnarray}
m_{\eta_8}^2=(m_\eta\cos\theta_P)^2+(m_{\eta^\prime}\sin\theta_P)^2
=(572.73~\MeV)^2~,
\label{meta8}
\\
m_{\eta_1}^2=(m_\eta\sin\theta_P)^2+(m_{\eta^\prime}\cos\theta_P)^2
=(943.05~\MeV)^2~,
\label{meta1}
\end{eqnarray}

\noindent where
$\theta_P=\phi_P-\arctan(\sqrt{2})=(-12.295\pm 0.019)^\circ$.
The mesons  $\eta_8$ and $\eta_1$ are defined as

\begin{eqnarray}
|\eta_8\rangle &=& \frac{1}{\sqrt{6}} (|u\bar{u}\rangle + |d\bar{d}\rangle - 2 |s\bar{s}\rangle)~, \\
|\eta_1\rangle &=& \frac{1}{\sqrt{3}} (|u\bar{u}\rangle + |d\bar{d}\rangle + |s\bar{s}\rangle)~.
\end{eqnarray}

\noindent The $\eta_8$ meson mass (\ref{meta8})
$m_{\eta_8}=572.73~\MeV$
is 4.56\% greater than the observed \cite{PDG2004}
$m_\eta=(547.75\pm 0.12)~\MeV$.
The singlet $\eta_1$ mass (\ref{meta1})
$m_{\eta_1}=943.06~\MeV$ is only
1.56\%  below the observed
$m_\eta^\prime=(957.78\pm 0.14)~\MeV$ and close to the 
nonstrange--$\bar{s}s$ mixing $U_A(1)$ mass dictated by phenomenology 
\cite{Jones:1979ez,Kekez:2000aw,Klabucar:2000me}
\begin{equation}
m_{U_A(1)}
\equiv (3\beta)^{1/2} =
\left[ \frac{3}{4} \frac{(m_{\eta^\prime}^2 - m_\pi^2) 
              (m_\eta^2 - m_\pi^2)} {m_K^2-m_\pi^2} \right]^{1/2}
= 915.31~\MeV~,
\label{secondU1Amass}
\end{equation}
(This is also close to $912~\MeV$, which is the mass found in the 
analogous DS approach \cite{Kekez:2000aw,Klabucar:2000me}.)

\begin{figure}
\centerline{\includegraphics[height=30mm,angle=0]{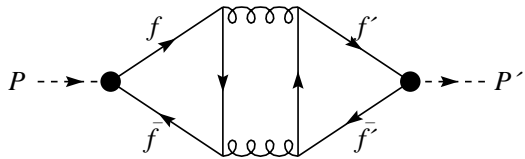}}
\caption{Nonperturbative QCD annihilation of a
quark-antiquark bound state illustrated by the diagram with
two-gluon exchange. The $\bar{q}q$ pseudoscalar $P$ is coupled
to a quark loop, whereby it can annihilate into gluons which in
turn recombine into the pseudoscalar $P'$ having the flavor
content $\bar{q'}q'$.}
\label{fig:diamond}
\end{figure}

We call the quantity (\ref{secondU1Amass}) the ``mixing $U_A(1)$ mass"
since the mass matrix (which is especially clear in the 
nonstrange-strange quark basis) reveals that $m_{U_A(1)}$ 
induces the mixing between the nonstrange isoscalar
$(|\bar{u}u\rangle+|\bar{d}d\rangle/\sqrt{2}$ and $\bar{s}s$
quark-antiquark states. Equivalently, $m_{U_A(1)}$ can be viewed 
as being generated by the transitions among the $\bar{u}u$,
$\bar{d}d$ and $\bar{s}s$ pseudoscalar states; via quark loops, 
these pseudoscalar $\bar{q}q$ bound states can annihilate into 
gluons which in turn via another quark loop can again recombine
into another pseudoscalar $\bar{q}'q'$ bound state of the same 
or different flavor. The quantity $\beta$ appearing in Eq.
(\ref{secondU1Amass}) is then the annihilation strength of such 
transitions, in the limit of an exact SU(3) flavor symmetry. 
(The realistic breaking of this symmetry is easily introduced 
and improves our description of the $\eta$-$\eta'$ complex 
considerably.)
The ``diamond'' graph in Fig.~\ref{fig:diamond} gives just the
simplest example of such an annihilation/recombination transition.
Since these annihilations occur in the nonperturbative regime of 
QCD, all graphs with any even number of gluons instead of just 
those two in Fig.~\ref{fig:diamond}, can be just as significant in annihilating 
and forming a $C^+$ pseudoscalar $\bar{q}q$ meson.  Indeed, this 
nonperturbative $U_A(1)$ mass scale, 
Eq.~(\ref{secondU1Amass}), is still 3 times higher than the 
gluon ``diamond'' graph evaluated perturbatively \cite{Choudhury}.
Thus, we cannot calculate $\beta = m_{U_A(1)}^2/3$ and the 
situation is much more complicated and less clear than in 
the Abelian case, as explained in \cite{Kekez:2005kx}.
Can it then be founded to think that the annihilation graphs with 
the pseudoscalar meson-quark coupling, such as the ``diamond'' 
graph in Fig.~\ref{fig:diamond}, give rise to the anomalous term 
$({3 \, g^2 }/{32 \pi^2})
\epsilon_{\mu\nu\alpha\beta} G_a^{\mu\nu} G_a^{\alpha\beta}$
in the divergence (\ref{divGanom}) of the SU(3)-singlet current 
$A^\mu_0(x)$, and thus ultimately to the large mass of $\eta_0$ 
(and of the observed $\eta'$)? Well, this conjecture may remain 
a speculation since we cannot calculate $\beta$ due to the
nonperturbative nature of the problem. Nevertheless, when
we use it in our approach as a parameter with the value given by 
Eq.~(\ref{secondU1Amass}), we obtain a very good description of 
the $\eta$-$\eta'$ complex phenomenology \cite{Jones:1979ez,Kekez:2000aw,Klabucar:1997zi,Klabucar:2000me,Klabucar:2001gr}.
This includes not only the masses of $\eta$ and $\eta'$, but also
their $\gamma\gamma$ decay widths, and the mixing angle
$\theta_P \approx -13^\circ$ consistently following from
the masses and $\gamma\gamma$ widths. 
This gives a strong motivation for the above conjecture.

\subsection{Taming of strong CP problem}

We should also note that our conjecture in the previous subsection
goes well with the arguments of Banerjee {\it et al.} 
\cite{Banerjee:2000qw}, that there is really no strong CP problem. They
find that one does {\it not} need vanishing 
$\Theta_{\rm eff} = \Theta - {\rm tr}\ln {\cal M}_q$ 
(where ${\cal M}_q$ is the quark mass matrix). Thus,
one does not need any fine-tuning, and all CP violation in 
the QCD Lagrangian can be avoided by having $\Theta = 0$ 
in its CP-violating term
\begin{equation}
{\cal L}_\Theta =
 - \Theta \, \frac{ g^2 }{64 \pi^2}
\epsilon_{\mu\nu\alpha\beta} G_a^{\mu\nu} G_a^{\alpha\beta}\, .
\label{ThetaQCD}
\end{equation}
This term in the QCD Lagrangian breaks the $U_A(1)$ symmetry
and corresponds to the anomalous term $\propto 
\epsilon_{\mu\nu\alpha\beta} G_a^{\mu\nu} G_a^{\alpha\beta}$
in the divergence (\ref{divGanom}) of the singlet current.
The term (\ref{ThetaQCD}) is allowed by gauge invariance and 
renormalizability, but apparent nonexistence of the strong 
CP violation, and also of axions, 
is the solid reason to have it vanishing. 
Our conjecture, that P-coupled annihilation graphs reproduce 
the effect of the gluon ABJ anomaly, naturally agrees with
the vanishing of this term and with putting the case of the
strong CP problem to rest {\it \`a la} Banerjee {\it et al.}
\cite{Banerjee:2000qw}.

\section{Summary/Discussion}
\label{sec:summary}

We have presented and surveyed in detail the method of 
pseudoscalar coupling of pseudoscalar mesons to the 
``triangle" and ``box" quark loops. We have reviewed
how this method gives the equivalent results to the 
anomaly calculations. The P-coupling method has also
been illustrated on the example of many decay amplitudes.

The AVV anomaly \cite{Adler69,BellJackiw69} involves 10 invariant
amplitudes (reduced to 1 or 2 amplitudes for $\pi^0\to \gamma\gamma$
decay using additional Ward identities). If instead one considers
the PVV transition with a pseudoscalar
coupling, then the PVV quark triangle amplitude is
finite and leads to many decay amplitudes (over 15)
then in agreement with data to within 3\% and not
involving free parameters \cite{Delbourgo}. To solve
instead the former AVV decay problem, very light axion
bosons have been predicted but have {\it not} yet been
observed \cite{PDG2004}.

   Also, there is the $U_A(1)$ and $\Theta$ problem involving gluons
whereby strong interaction QCD leads to CP violation,
definitely a ``strong CP problem'' because CP violation is known
to occur at the $10^{-3}$ weak interaction amplitude level \cite{PDG2004}.
Physicists have tried to circumvent this ``$U_A(1)$  --
strong CP problem'' either via the topology of gauge
fields or by investigating the $\Theta$--vacuum for this
strong CP problem \cite{Banerjee:2000qw}.

   In this paper we have circumvented the need to deal directly
with the above photon or gluon AVV anomalies by studying 
instead (finite) PVV quark triangle graphs. Then we have given 
our phenomenological results -- which always are in
approximate agreement with the data. Next we return to
the $U_A(1)$ problem and again use quark triangle
diagrams coupled to 2 gluons. Invoking nonstrange--strange
particle mixing, the predicted $U_A(1)$
mass is within 3\% of data
\cite{Jones:1979ez,Kekez:2000aw,Klabucar:2000me}.

   Thus we circumvent both photon and, admittedly on a
much more speculative level, also the gluon ABJ anomaly
without resorting either to unmeasured axions or to
a strong CP violating term in the QCD Lagrangian.


\end{document}